# Analysis of the trends in the index of the Dow Jones Industrial Average (DJIA) of the New York Stock Exchange (NYSE)

Çağlar Tuncay
caglart@metu.edu.tr

**Abstract:**

It is hypothesized that price charts can be empirically decomposed into two components as random and non random. The non random component, which can be treated as approximately regular behavior of the prices (trend) in an epoch, is a geometric line. Thus, the random component fluctuates around the non random component with various amplitudes. Moreover, the shape of a trend in an epoch may be different in another epoch. It is further hypothesized that statistical evidence can be found for various relations between several types of trends and the direction of the next movements of the prices.

These hypotheses are tested on the historical data of the DJIA (Dow) and confirmed. Moreover, it is statistically showed that a number of trends that have occurred in the near past course of the Dow can be utilized to presage the near future of the index. As a result, upcoming of a recession in the DJIA, which may portend a worldwide economic crisis, is predicted.

## 1. Introduction:

It is known that the DJIA is the oldest and most voluminous index of the NYSE which is the oldest stock exchange. Therefore, the Dow can be regarded as a relevant representative of world market. Hence, this index is worth studying with the aim of understanding the past, present and especially, forecasting the future of world economy.

Thankfully, a number of quantitative methods have been proposed and used in the literature for statistical finance, also called econophysics, with the aim of investigating various aspects of price charts [1]. Yet, the capability of those methods to forecast the future prices is not verified scientifically. Moreover, no academic consensus is established on the answer to the question whether series of speculative prices can be forecasted, which has been an important issue in the agenda of the scholars for about a century. The reader is referred to the following papers [2] and books [3] for comprehensive studies on a number of econophysical methods, such as the efficient market theory and technical analysis. It should be noted that a number of known methods and new ones will be used in this work.

**1.a Model**: This work deals with the time (t) variation of the adjusted (cum dividend) daily close values ($\chi(t)$) of the DJIA. It is well known that the $\chi(t)$ is calculated in terms of weighted average of several asset prices in the NYSE. However, the inflation rate is not subtracted from the Dow in this work.

***<u>Epochs and trends</u>***: The basic assumption of the model is that the DJIA has followed a simple trend for each of several epochs, plus random fluctuations. Moreover, these trends can be used to forecast the near future prices up to random fluctuations. Furthermore, a simple trend can be described by a straight line, exponential line, parabola or sinus cycle. It is worth noting that the parabolic and cyclic trends are used to be ignored by a number of technical analysts [2] and [3]. However, also the sinus cyclic and especially parabolic behaviors are found to have important clues for the next movements of the prices, as will be described in detail at the end of this section and exemplified in further sections.

It should be noted that the time interval which starts and ends when a mid-range trend (MRT) starts and ends is called epoch in this work. Moreover, durations of the MRTs are a number of months or years. Furthermore, an epoch may consist of several sub-epochs which cover shorter trends. A number of basic epochs and their sub-epochs in the historical data of the Dow will be studied in Section 2.b.

The following technique is proposed for the identification of the MRTs. The locally extreme (maximum or minimum) values of the prices can be connected by two lines and then, these lines can be segmented into straight lines, exponential lines, parabolas or sinus cycles. It is worth noting that the line connecting the locally maximum (minimum) values are called resisting (supporting) line in the literature [2] and [3]. Moreover, the resisting and supporting lines may be parallel in a given epoch where the result is a band. In this case, the average of the resisting and supporting lines can be taken as the trend characterizing the given epoch. It is obvious that least square fittings (LSFs) can be further applied on the lines obtained by the current technique.

It is worth noting that when the supporting or resisting line for a given trend is crossed down or up by the prices, then this means that the exploited trend ended and the next one will form below or above the exploited prices, respectively. This defines how the basic epochs are distinguished in Dow (see, Section 2.b).

Another technique may be applying LSF directly on a selected segment of the raw data for one of the expected trends where the coefficient of determination ($R^2$) is known to be a measure for the success of the fit. To be more precise, if $R^2$ for a least square fit is close to zero (unity), then the fit is concerned as a bad (good) one. The reader is referred to any standard book on the probability theory or a paper such as [4] for further reading about fits and $R^2$.

The model can be described more precisely in the following manner. Suppose that the integer ($0<n$) enumerates the MRTs in chronological order and the $\tau_n$ is the time with regard to the beginning date of the $n^{th}$ MRT; i.e., $\tau_n=0$ at the date when the previous, i.e., $(n-1)^{th}$ MRT ended or $\tau_1= t =1$. Then, the prices $\chi(t)$ in the $n^{th}$ MRT can be written as

$$\chi(t) = X_n(\tau_n) + \varepsilon(\tau_n) \tag{1}$$

where the $X_n(\tau_n)$ stands for the $n^{th}$ MRT and the $\varepsilon(\tau_n)$ represents the random component. As a result, the daily returns ($\Delta\chi(t)$)

$$\Delta\chi(t) = \chi(t) - \chi(t-1) \tag{2}$$

can be written as

$$\Delta\chi(t) = X_n(\tau_n) - X_n(\tau_n-1) + (\varepsilon(\tau_n) - \varepsilon(\tau_n-1)) \tag{3}$$

or $$\Delta\chi(t) = \Delta X_n(\tau_n) + \varepsilon'(\tau_n) \tag{4}$$

where the $\Delta X_n(\tau_n)$ ($= X_n(\tau_n) - X_n(\tau_n-1)$) stands for the daily returns in the $n^{th}$ MRT apart from the random fluctuations and $\varepsilon'(\tau_n)$ represents a random series since the successive time differences of a random time series is another random time series. It is obvious that the $\Delta X_n(\tau_n)$ will attain a constant value in a linear up or down trend and it will exponentially raise or fall in an exponentially up or down trend, respectively. Moreover, the $\Delta X_n(\tau_n)$ will alternate with the time in parabolic and cyclic trends.

It is obvious that the character of the $\Delta X_n(\tau_n)$ is expected to change from one trend to the other. It is worth noting that also the rate of change of the daily returns ($\Delta\Delta X_n(\tau_n)$) which can be written simply as

$$\Delta\Delta X_n(\tau_n) = \Delta(\Delta X_n(\tau_n)) = \Delta X_n(\tau_n) - \Delta X_n(\tau_n\text{-}1) \quad (5)$$

is characteristic for a trend.

***Speeds and accelerations***: More precisely, the daily returns can be treated as the daily speeds (v(t)) of the price movements but only numerically since

$$v(t) = (\chi(t) - \chi(t\text{-}1))/(t\text{-}(t\text{-}1)) = \Delta\chi(t) \quad (6)$$

or $$V_n(\tau_n) = (X_n(\tau_n) - X_n(\tau_n\text{-}1))/(\tau_n\text{-}(\tau_n\text{-}1)) = \Delta X(\tau_n). \quad (7)$$

Moreover, the rates of changes of the $\Delta\chi(t)$ or $\Delta X_n(\tau_n)$ can be treated as accelerations of the price movements, but only numerically;

$$a(t) = (v(t) - v(t\text{-}1))/(t\text{-}(t\text{-}1)) = \Delta v(t) \quad (8)$$

or $$A_n(\tau_n) = (V_n(\tau_n) - V_n(\tau_n\text{-}1))/(\tau_n\text{-}(\tau_n\text{-}1)) = \Delta V_n(\tau_n). \quad (9)$$

It is obvious that the $A_n(\tau_n)$ will attain a value around zero in a linear trend and it will vary in proportion to the $X_n(\tau_n)$ in exponential and sinus cyclic trends. Moreover, the $A_n(\tau_n)$ will attain a constant positive (negative) value in an open up (open down) parabolic trend. The reader is referred to Table 1 for the time equations, and the speeds and accelerations of the price movements in the studied MRTs.

Conclusively, the character of the $A_n(\tau_n)$ as well as that of the $V_n(\tau_n)$ is expected to change from one trend to the consecutive trend.

***Formations***: Several complicated trends which consist of simple ones are important for understanding several regimes in the Dow since they have occurred frequently in the price chart. These complicated trends will be called formations henceforth. Moreover, it is a commonly shared opinion among the chartists [2], [3] and [5] who claim that various formations such as, the shoulder-head-shoulder (SHS) and reversed shoulder-head-shoulder (RSHS) formations indicate important falls and jumps, respectively. If the prices are higher in an epoch with a given MRT with regard to those in the previous and latter epochs with similar or different MRTs, then this formation is called SHS. If, on the other hand, the prices are lower in an epoch with a given MRT with regard to those in the previous and latter epochs with similar or different MRTs, then this formation is called reversed SHS (RSHS). It is obvious that a SHS (RSHS) formation can be interpreted as a long lasting open down (open up) parabola with regard to the individual MRTs where the three local maximum (minimum) prices of the individual three MRTs can be used for defining the large parabolas. This explains the reason for the occurrence of important falls and jumps after the occurrence of SHS and RSHS formations, respectively. More precisely, the SHS or RSHS formations define the top or bottom of large open down or open up parabolic MRTs and thus, the prices show long lasting downward or upward accelerations, respectively. Moreover, SHS (RSHS) formations may occur in terms of three consecutive open down (open up) parabolas. As a result, the prices display constant downward (upward) accelerations with long durations within SHS (RSHS) formations; see the equations for parabolic trend in Table 1. Furthermore, the line connecting the local minimum values of the head in a SHS formation or that connecting the local maximum values of the head in a RSHS formation can be taken as the maturation level of those formations. More precisely, when the prices, following the right shoulder in either the SHS or RSHS formations, cross down or up the previously defined line, then the prices can be expected to further fall or rise, respectively. On the other hand, several other formations, such as double tops (bottoms) can also be described in terms of two consecutive open down (open up) parabolic behaviors. Therefore, parabolic MRTs should not be omitted in trend analyses.

**1.b Material**: The daily raw data used in this work have been downloaded from the URL given in [6].

The results of the analysis of the past and forecasts of the near future of the DJIA are presented in the following section. The last section is devoted to further discussion and conclusion.

**2. Analysis of the Dow**:

The daily time chart of the DJIA with a time domain from the establishment day at 01 Dec 1928 (t=0) until 02 Sep 2011 (t=20,825), inclusively and the corresponding daily trading volume are showed in the upper and lower parts of Figure 1 with logarithmic vertical axes on left and right, respectively. It should be noted that similar axes will be used in Figures 3 – 7. Moreover, the capital in dollar exchanged per day in the Dow, except various expenses, such as the commissions, fees, taxes, is called daily trading volume of the Dow in this work; see [6].

**2.a An overall look:** On inspection of Fig.1, which covers 20,825 trading days between 1 Now 1928 and 2 Sep 2011, it can be observed that the overall behavior of the Dow can be investigated in five main epochs. Actually, several short-term time intervals with unimportant or no trends have occurred between the successive epochs with the MRTs, which are disregarded in this work.

Moreover, the exponential growth can be observed in the both price and trading volume of the Dow. For example, the volume has decreased exponentially on average, from the establishment day of the index until around the year 1945. Afterwards, the volume has ascended exponentially until the present. It may be that these exponential behaviors are due to the exponentially growing economy of the country. In order to test the validity of this hypothesis, the time course of the gross domestic product (GDP) index of the USA which is a measure of country's overall official economic output[7] and that of the Dow are plotted in Figure 2 with logarithmic vertical axes on the right and left, respectively. As with the Dow, the inflation rate is not subtracted from the shown GDP.

On inspection of Fig. 2 it can be observed that the time domains of the GDP and DJIA indices are not common. This is because the GDP statistics are available only quarterly, and for the years between 1947 and 2011, exclusively. Moreover, both indices have grown exponentially; however, with different exponents. It is obvious that the relative increase of Dow is larger than that of the GDP index as the dashed line, for the Dow, is steeper with regard to the dotted line, for the GDP index, in Fig.2. It is noteworthy that the corresponding $R^2$ terms which measure the goodness of the fits are 0.931 and 0.995 for the Dow and GDP, respectively. Furthermore, the fluctuations in the GDP index are small with regard to those in the DJIA. Therefore, these indices may be correlated but weakly. This means that USA economy may be decisive for the pattern of the Dow in the long-term; however, it does not define the underlying mid-term or short-term dynamics of the Dow. In other words, the exponential behavior of the GDP index is steady whilst the horizontal and exponential characters of the Dow alternate, as will be discussed in detail in Section 2.b).

**2.b Trend analysis**:

The history of Dow can be investigated in five basic epochs; see, Fig. 1. The criterion used for distinguishing these epochs has been discussed under the heading Epochs and trends in Section 1.a. Moreover, the important lines ($L_i$) and prices ($P_j$) will be designated by dashed (and straight) and dotted (and curled) arrows in the figures of this section.

**EPOCH 1**: This epoch begins with the establishment of the index on 01 Dec 1928 ($\chi$=$239.43) and lasts until around 04 Jan 1943 ($\chi$=$120.25) with duration of around 3,500 trading days, and it consists of three sub-epochs with strong exponential decrease, exponential increase, and weak exponential increase; see Figure 3.

***Sub-epoch with the strong exponential decrease***: This sub-epoch covers the Great Depression (Wall Street Crash of 1929, also known as the October 1929 crash) which lasted around 700 calendar days. The climax of the sub-epoch has occurred at an elevation of $381.17 ($P_1$ in Fig. 3) and the locally minimum price level has occurred at value of $41.22 which is the overall lowest value of the Dow ($P_2$ in Fig. 3).

The technical question here is how the end of the recession could have been identified before long time passed over it. If this point had been identified timely, then acquisitions could have been made safely. The answer to this critical question is waiting for the occurrence of a new local minimum above the previous one; and thus, a supporting line could have been defined for the near future prices. It can be observed in Fig. 3 that the $L_1$ line connecting the two closely located local minimum prices ($P_2$ and $P_3$ in Fig. 3) has served as a supporting line. What if the prices had crossed the $L_1$ line and fallen below the second locally minimum price? In this case, the acquiree would have been sold and new acquisition could have been made right after the prices go over the recent local minimum. However, it is not certain that there would be no further local minima. For example, several successive local minima have occurred during the October 1929 crash. In other words, if a local minimum would have occurred above the one dated to the middle of 1932 and a third one would have occurred below the first local dip, then this would have implied that the effect of the October 1929 crisis on the crash had not ended yet. Or vice versa, if the second local dip had occurred above the first one, and the third one, if any, has occurred above the second local dip then, this would have implied that the effect of the October 1929 crash on Dow has ended.

***Sub-epoch with the exponential increase***: On inspection of Fig. 3 it can be observed that after a short-range transition term, the index movements have changed direction and an upward exponential bullish trend has started around the middle of 1932. Afterwards, the index has ascended to $194.40 within few years ($P_4$ in Fig. 3). It is noteworthy that the DJIA has not fallen below the line denoted by $L_1$ in Fig.3, which connects two locally minimum values ($P_2$ and $P_3$ in Fig. 3); and hence, this line has functioned as a supporting line in the sub-epoch. Astonishingly, the behavior of the Dow in the following nearly five years, has been defined by the above mentioned two extreme values ($P_2$ and $P_3$ in Fig. 3). More precisely, the index has not crossed down the line denoted by $L_1$ in Fig.3 until around the middle of 1937.

***Sub-epoch with the weak exponential increase***: The prices in the next sub-epoch; i.e., between the days for the prices $P_5$ and $P_7$ in Fig. 3, have displayed exponentially decreasing behavior with the exponent equaling approximately to the slope of the line $L_2$ in logarithmic scale. It is noteworthy that the line $L_2$ connects the local climax at $381.17 ($P_1$ in Fig. 3) and the locally highest prices at $192.91 at around the beginning of 1937 ($P_4$ in Fig. 3) and those at around the end of 1943 ($P_7$ in Fig. 3), and a number of them in between. Hence the duration of functioning of this line to resist against the prices in the sub-epoch is around six calendar years. Moreover, it has functioned as a resisting line not only in this sub-epoch but also through out the EPOCH 1. In other words, the line $L_2$ is decisive for the identification of the EPOCH 1. This is because the prices have remained below that line for around twelve calendar years between 1929 and 1943. Moreover, the prices rose after crossing this line over ($P_7$ in Fig. 3) without further testing it. This event can be concerned as providing statistical evidence for the verification of a commonly shared opinion among the traders who claim that the prices follow the direction which has occurred right after the ending of the previous trend [2], [3] and [5]. On the other hand, the interesting point is that the behavior of the index between 1937 and 1943 has been controlled by two previous events during which the index has formed the two locally extreme values; see $P_1$ and $P_4$ in Fig. 3. In other words, the near past and present has affected the near future of the Dow. This issue will be further discussed under the next headings and in the last section.

On the other hand, the index has experienced upward or downward accelerations when it tested the supporting or resisting lines several times. More precisely, the downward speeds have changed to upward speeds and vice versa, during those tests. Moreover, the index has experienced an upward acceleration with large magnitude when the index has turned upward after the Wall Street Crash of 1929 ($P_2$ in Fig. 3). However, the DJIA has experienced downward acceleration at around the end of 1936 as the index has tested the resisting line ($L_2$) the second time (between $P_4$ and $P_5$ in Fig. 3). It may be expected that voluminous trading accompanies have applied large upward accelerations. However, this expectation is not verified statistically. For example, the upward acceleration has occurred with small daily volumes, at around the middle of 1942 ($P_6$ in Fig. 3). The reader is referred to [8] for the role of volume in the price movements. It is worth noting that several mechanisms used to accelerate the index will be discussed and exemplified under further headings.

It is claimed that the main characteristic of this epoch is squeezing the prices exponentially as designated by two arrows $L_2$ and $L_3$ in Fig.3. The arrow $L_3$ in Fig.3 connects two local minimum values one of which is the turning value of the November 1929 crash and the other is the local minimum of the recently mentioned time term when acceleration is applied on the index with small volume at around the middle of 1942; $P_2$ and $P_6$ in Fig. 3, respectively. These two local minima have occurred at the elevations \$41.22 and \$92.92 during the 946$^{th}$ and 3390$^{th}$ days after the establishment of Dow, respectively. Therefore, the exponent ($\Omega_3$) and hence the equation of this line can be obtained as follows.

$$\Omega_3 = \ln(92.92/41.22)/(3390-946) = 0.000333 \text{ day}^{-1}$$

and $$L_3(\tau) = 41.22\exp(\Omega_3(\tau-946)) = 92.92\exp(\Omega_3(\tau-3390)) \quad . \qquad (10)$$

It should be noted that, $\tau$ is the same as t in this case, since this is the first basic epoch. Moreover, this epoch has ended when the prices have crossed the line $L_2$ over at around 04 Jan 1943.

In summary, EPOCH 1 can be identified in terms of two exponential lines ($L_2$ and $L_3$ in Fig.3) which met each other in late 1943. Thus, these lines have functioned as supporting or resisting lines and hence the prices have been squeezed. Moreover, the line $L_2$ is decisive for the identification of the line $L_3$ and thus that of the EPOCH 1. This is because the line $L_2$ has connected a number of locally maximum prices until around the end of 1942 and the prices have gone over this line at 04 Jan 1943. Therefore, the line $L_2$ is well defined. Once the line $L_2$ is identified, then a second line called $L_3$ can be identified as connecting the locally minimum prices which would have occurred before 04 Jan 1943. It can be observed in Fig. 3 that $P_2$ and $P_6$ are such prices; moreover, a line connecting them can be taken as the supporting line of EPOCH 1. As a result, the characteristic of this epoch is squeezing the prices which have converged to around \$100 as can be observed in Fig. 3.

**EPOCH 2**: The daily time course of the Dow with volume during the second basic epoch is displayed in Figure 4. This epoch has begun at around 04 Jan 1943 ($P_7$ in Figs. 3 and 4) and continued until 13 Feb 1964 (the 8854$^{th}$ day with $\chi$=\$794.42) ($P_9$ in Fig. 4). Hence, the duration of this basic epoch, with several unimportant sub-epochs, is approximately 5,474 trading days.

On inspection of Figure 4 it can be observed that this epoch has started when the prices have crossed over the line $L_2$ (in Figs. 3 and 4) at the beginning of 1943 and ended at the beginning of 1968 ($P_9$ in Fig. 4) when the prices fell below the line $L_3$ (in Fig. 4). Hence, the line $L_3$ from EPOCH 1 has continued to function as supporting the prices in this epoch, as well. As a result, the characteristic of this epoch has been exponential growth with the exponent approximately equaling to that of the line $L_3$; see, Eq. 10.

On the other hand, the growth has shown temporarily steeper character several times in the epoch and the line passing through the local maximum values can be taken as the resisting line of the exponential MRT ($L'_3$ in Fig. 4). Thus, an exponential band can be formed. However, the increase has started to lose speed when the prices have approached to the level of \$1000 (around $P_8$ in Fig. 4). Moreover, the prices have tested this level several times later. Therefore, the horizontal line at the level of \$1000 ($L_4$ in Fig. 4) has functioned as another resisting line in this epoch. Finally, the prices have crossed down the line $L_3$ in Fig. 4 at around $P_9$ in Fig. 4 and the MRT has ended. It is noteworthy that the time term between the days for $P_8$ in Fig. 4 and $P_9$ in Fig. 4 can be taken as transition between this and the next basic MRT.

An interesting feature of this epoch is that the prices have tested the line $L_3$ in Fig. 4 several times from above without crossing it down until the beginning of 1969 ($P_9$ in Fig. 4). As a result, the duration of this line for supporting the prices is around 37 calendar years; more precisely, the concerned mood has continued from around the end of 1932 in EPOCH 1, until around the beginning of 1969. On the other hand, several objections can be posed for that feature of the Dow; for example, it may well be argued that this is accidental. But, the point here is that a different regime which will be studied under the next heading has started after the prices have crossed down the line $L_3$ in Fig. 4 at around the beginning of 1969 ($P_9$ in Fig. 4) as this MRT had started after the prices have crossed over the resisting line $L_2$ in Figs. 3 and 4 in the previous MRT; see $P_7$ in Figs. 3 and 4. Therefore, the line $L_3$ has supported the prices in two different regimes, which may not be accidental. Moreover, the line $L_3$ has shown various more interesting behaviors within the next coming MRTs as will be described under further headings.

**EPOCH 3**: The third epoch has been started on around 13 Feb 1964, including the previously mentioned transition term and continued till around 04 Jan 1982 (13,340$^{th}$ day with $\chi$=\$882.52). Hence, the duration of this epoch is around 4,488 trading days or 18 calendar years. The characteristic of this epoch is the horizontal behavior of the metrics around the level of \$1000; more precisely, in a channel between nearly \$750 and \$1000. In other words, the lines for the upper and lower bounds are horizontal; see Figure 5 where the vertical axis on left for Dow is linear and that on right for the volume is logarithmic. The reader may be referred to [2], [3] and [5] for technical analyses of channels in price charts. On the other hand, the daily volume increases exponentially on average in this epoch. This shows statistically that the prices and volumes may be independent, as underlined in several sections.

An interesting feature of this epoch is the nearly cyclic behavior of the index with a period of 750 trading days, if various deformations are disregarded. The sinusoidal fit which is shown by a thick line in Fig. 5 has the following equation:

$$X(\tau_3) = 875 + 125\sin(\tau_3(2\pi/750) - \pi/2) \quad (11)$$

where $\tau_3 = 0$ on 21 Apr 1964 which is the 8,900$^{th}$ day of the Dow. It is worth noting that the day for $\tau_3 = 0$ is not the day when the previous epoch has ended, which is the 8854$^{th}$ day with $\chi$=\$794.42 (13 Feb 1964); see, $P_9$ in Figs. 4 and 5. Therefore the day for the date 21 Apr 1964 can be taken as the beginning of the transition time term between the previous epoch and this one. Moreover, $R^2$ of the sinus LSF to Dow in the time domain between 21 Apr 1964 and 01 Jan 1982 is approximately 0.89 which indicates that the fit is not bad. Such approximately cyclic or harmonic oscillations occur frequently in the price charts as it will be further discussed under the last heading in this section.

On inspection of Fig. 5 it can be observed that the line $L_3$ is exponential since the vertical axis for Dow is linear. Moreover, the two tops and one bottom have occurred before the price $P_9$. Hence, these two tops could have been connected by a line for the resisting line and a parallel one to that line and passing from the dip could have been drawn as a supporting line. Another decision would be waiting for the clear occurrences of two tips and two dips to obtain more reliable lines. Once these lines are identified, then acquisitions could have been made when the prices fall to around \$750 with the aim of selling them when the prices ascend to around \$1000 later. However, the prices have crossed down the supporting line twice in this epoch; more precisely, at around the beginning of 1970 and at around the beginning of 1975.

**EPOCH 4**: The previous horizontal MRT has ended with an abrupt rise of the index at around the beginning of 1982, and later, the index has exponentially ascended in this epoch as can be observed in Figure 6. This event can be concerned as providing statistical evidence for the verification of a commonly shared opinion among the traders who claim that the prices follow the direction which has occurred right after the ending of the previous trend [2], [3] and [5].

Actually, this epoch has consisted of two sub-epochs each displaying exponential growth with different approximate exponents of 0.0005 per trading day between 01.01.1982 and 08.12.1994, and 0.0008 per trading day afterwards, as designated by the arrows $L_4$ and $L_5$ in Fig. 6, respectively. It is worth noting that the index had crossed down the line $L_3$ at around the beginning of the previous epoch. On inspection of Figures 6 and 7 it can be observed that the index has approached from bottom to the line $L_3$ in 1999 and at the beginning of 2000 due to the above mentioned two sub-epochs for rapid exponential growths. However, the index could not have crossed it over even though several tests have been applied. As a result, the index has started to fall at around the beginning of 2000 and thus, the fourth MRT has ended. Therefore, the duration of this epoch has been around 4,660 trading days or around 18 calendar years.

An interesting feature of this epoch is that the line $L_3$ which had been identified as a supporting line in terms of the two local minimum prices of the Dow in around November 1929 and around the middle of 1942 behaved as a resisting line in this MRT. Therefore, the importance of the $L_3$ line in the history of Dow has continued during the last 70 calendar years. However, no claim will be made about the answer to the question whether such long lasting importance of the $L_3$ line is due to the planned behavior of the traders or underlying dynamics of the Dow, which is not known at the present. The reader is referred to the description around Eq. 10 under the heading of EPOCH 1 in Section 2.b, for the identification of the $L_3$ line. Moreover, the overall behavior of the $L_3$ line can be taken as providing statistical evidence for the verification of a commonly shared opinion among the traders who claim that a supporting (resisting) line functions as a resisting (supporting) line after it has been crossed down (over) by the prices [2], [3] and [5]. Furthermore, a number of traders claim that a supporting (resisting) line cannot be crossed down (over) without several tests from above (below), prior to the turning point [3].

**EPOCH 5**: Behavior of the DJIA index and the daily volume in the current epoch is displayed in Figure 8 with linear axes. On inspection of Fig. 8 it can be observed that the characteristic of this epoch is oscillations around \$10,000, which can be described in terms of six overlapping or separate parabolas as designated by $Q_i$ with $1 \le i \le 6$ in the figure. Only the parabolas $Q_1$, $Q_2$, $Q_4$ and $Q_6$ will be studied here.

***Sub-epoch with the parabola $Q_1$***: The index showed horizontal behavior for several months after it had tested the $L_3$ line from bottom several times in 1999 and 2000. Later, it has descended with oscillations around the parabola $Q_1$ which is obtained by a LSF, as can be seen in Figure 9.

The equation of this parabola is

$$Q_1(\tau'_5) = 8796 + 25.2343\tau'_5 - 0.1145(\tau'_5)^2 \quad (12)$$

where $R^2 = 0.85$ and $\tau'_5=1$ is selected on 21.09.2001.

It is noteworthy that a SHS formation has occurred towards the end of this parabola. The left shoulder, head and right shoulder of the SHS formation are designated by $S^L_1$, $H_1$ and $S^R_1$ in Fig.6, respectively.

***Sub-epoch with the parabola $Q_2$***: Another SHS formation has occurred right after the previous SHS formation had been matured at around the end of 2002 and the beginning of 2003 as can be seen in Figure 10. The local minimum values of those consecutive SHS formations can be identified as the left shoulder, head and right shoulder of a reversed SHS formation, which are designated by $S^L_2$, $H_2$ and $S^R_2$ in Fig.10, respectively. Obviously, a parabola which is designated as $Q_2$ in Fig.10 can be defined in terms of those locally three minimum values.

The time trajectory of the parabola $Q_2$ follows the equation

$$Q_2(\tau''_5) = 7713 - 11.0965\tau''_5 + 0.0620(\tau''_5)^2 \quad (13)$$

where $\tau''_5=0$ taken on 23.07.2002.

It should be noted that the behavior of the prices during the above mentioned time intervals for the SHS or reversed SHS formations can be taken as providing statistical evidence for the verification of a commonly shared opinion among the traders who claim that the prices descend (ascend) after the formation of (reversed) SHS formations [3] and [5]. This is because open up (open down) parabolas may occur after the occurrence of reversed SHS (SHS) formations and constant upward (downward) accelerations are applied during open up (open down) parabolas. In other words, SHS or reversed SHS formations may be part of the larger open down or open up parabolas with long durations. As a result, acquisitions could have been made after the index crossed over the $L_6$ line in Fig. 10 at around \$9,000 with the aim of selling them at around the top of the parabola $Q_1$ which is higher than the \$10,000 level. It would have been expected that this strategy would provide around 10% gains within six to eight months. Actually, the index has ascended further along a parabola designated as $Q_3$ in Fig.9, later. Thus, the climax of all times of the Dow has occurred at the level of $14,160 which was attained by the index on 09 Dec 2007.

It is interesting that a cyclic behavior with a period of around 53 trading days and a downward average speed of -2.5 dollar per trading day has started at the end of 2003 and lasted for around a year, during the sub-epoch for $Q_3$; see Fig. 8. This cyclic segment of the Dow and the sinus fit are showed in Figure 11. The equation of the fitting line is

$$X(\tau'''_5) = 10500 - 2.5\tau'''5 + 300\sin(\tau'''5(2\pi/53)-\pi/2) \quad (14)$$

where $(\tau'''_5)=0$ is taken on 01.12.2003.

It is worth noting that this behavior could have been used for around 5% profits in around 28 trading days, which equals to half of the period, for each cycle but only after the occurrence of the second or more safely the third testing of the supporting $L_7$ line by the index. Moreover, the index has crossed over the resisting $L_8$ line and has continued to elevate towards the climax, as can be observed in Fig. 11.

Actually, a recession in the NYSE had been forecasted using trend analysis in a paper published at the beginning of 2006 [9] where the claim was that the DJIA will recede back to the 8000's in about 500 days following 23 June 2005 which was the received date of the paper. However, this claim was revised as "Perhaps these fluctuations signal a transition to a different regime, to be seen in the coming years" in [10]. The Dow has risen and formed the climax after those papers have been published and fallen to test the prices lower than $7000 in March 2009. As a result, that forecast can be taken as realized but with some delay.

***Sub-epoch with the parabola $Q_4$***: The behavior of Dow around the historical extreme prices is displayed in Figure 12 with linear axes, where the dotted line designates the parabolic fitting line $Q_4$. The equation of this line can be given as

$$Q_4(\tau''''_5) = 10564 + 21.7763\tau''''_5 - 0.0405(\tau''''_5)^2 \quad (15)$$

where $R^2$=0.93 and $\tau''''_5$=1 is selected on 19.07.2006. Therefore the Dow has experienced constant downward acceleration that is equal to (0.0405*2=) 0.08 $/day$^2$ on average, in this sub-epoch. As a result, the upward speed which was approximately equaling to 21.8 $/day as obtained from Eq. (15) at around 19.07.2006 has decreased with time and the behavior of the Dow has turned to be horizontal at around the day for the historical climax. Afterwards, a recession has occurred in which the index has fallen to $7000 in March 2009 from $14,160 which had been attained by the index in 09 Dec 2007. This crisis is known to have happened around the Lehman Brothers bankruptcy.

***Sub-epoch with the parabola $Q_6$***: The initiation of the recovery from the 2008 crisis with bankruptcies is known to be the period when high profits were declared to be made by the City Group in the first quarter of 2009. However, a reversed SHS formation which is denoted by the parabola $Q_5$ in Fig. 8 had been forming beforehand. More precisely, the head of the parabola $Q_5$ occurred in Mar 2008 and the right shoulder occurred in June of the same year and thus, the index rose above the $10,000 level linearly under the effect of the upward acceleration applied by means of the parabola $Q_5$ as can be observed in Fig. 8. In other words, the speed of the Dow has turned from downward to upward during the sub-epoch for the parabola $Q_5$ when the index was around $5000 in around the beginning of 2009 as can be seen in Figure 12. However, a linear bullish trend which is designated by the line $L_{10}$ in Fig. 13 started in Feb 2010 and ended at around the beginning of May 2010 at the elevation of around $11,000. It is worth noting that voluminous sales have caused the alteration of that linear up trend as can be seen in the lower line with the linear vertical axis on the right. Afterwards, an

open up parabolic trend has started to occur, which is designated by the line $Q_6$ in Fig.13. The equation for this line is

$$Q_6(\tau''''_5) = 10335 - 30.159\tau''''_5 + 0.3346(\tau''''_5)^2 \quad (16)$$

where $R^2 = 0.96$ and $\tau''''_5 = 1$ is selected on 10.05.2010. It is noteworthy that the line $Q_6$ is obtained by means of the following method instead of applying a parabolic LSF. A parametric equation quadratic in t, such as the one in Table 1, is solved using 3 prices which are \$10380.43 on 07.05.2010, \$10068.01 on 20.05.2010 and \$9974.45 on 26.05.2010. Moreover, those three days are prior to the day for the local minimum of the line $Q_6$. Therefore, a parabolic trend can be predicted prior to the maturation and suitable strategic decisions can be made for trading. Here, for example, it might have been expected that the index will rise after the completion of the open up parabolic formation due to the acceleration applied upward in the meantime. The level for the maturation of that parabola can be taken as the level when lines $L_{11}$ and $L_{12}$ crossed. The index has risen linearly, after the completion of the parabola $Q_6$, with a speed approximately equal to that of the linear bullish trend designated by the line $L_{10}$ in Fig.13. It is noteworthy that the index has departed from the line $L_{13}$ considerably in 2011; however, the line $L_{11}$ which had been functioning as the resisting line in the sub-epoch for the parabola $Q_6$, has worked to support the index in March 2011. As a result, this event provides statistical evidence for the commonly shared opinion among the chartists who claim that a supporting line may function as a resisting line later or vice versa [2], [3] and [5]. It should be noted that the line $L_3$ which resisted against the prices in 1999 and 2000 had worked as a supporting line for several years beforehand. Conclusively, the acquisitions made at a level around \$9700 in June 2010 should have been held until the prices departed from the line $L_{13}$ at a level around \$12,000 which amounts to ((12000-9700)/9700 = ) 23.7% gain within around 10 months.

***<u>Forecast</u>***: An important aspect of the current epoch is the possibility of occurrence of a long-term SHS formation. It seems that the prices close to the line $L_3$ in 2008 and those in the beginning of 2009 at a level of around \$12,000, and those around the historical climax of \$14,160 at 09 Dec 2007 can be taken as the left shoulder ($S_L$) and head (H) of that potential SHS formation, respectively. Therefore, the behavior of the Dow at the time around Sep 2011 is critical. If the prices do not ascend afterwards considerably, which is weakly possible due to the recent news about the weak rates of USA and global economy, then, the current behavior may be taken as the right shoulder ($S_H$) the local climax of which has occurred on 29 Apr 2011 at \$12810.54; see Fig. 14. On inspection of Fig. 14 it may be observed that the recently mentioned three local maximum prices can be used to define an enveloping parabola for the expected SHS formation and this envelope covers the prices within a time domain starting around 1970 until 02 Sep 2011 or longer. Moreover, the associated downward acceleration is around 0.0005 \$/day$^2$; see the parameter ($\Lambda$) in Table 1.

In case of the occurrence of a long-term SHS formation, it is expected that the prices will fall, under the effect of the long-lasting downward acceleration ($\Lambda$=0.0005 \$/day$^2$) due to the expected SHS formation and the enveloping parabola, until the level of \$4,000 or lower. The technical reasons for this prediction can be summarized as follows.

1- The maturation level of the prospective formation is expected to take place when the future prices cross down the line connecting the local minimum values in the open up parabolas $Q_2$ and $Q_5$ in Fig. 8 as shown as the line $L_{16}$ in Figure 14. On inspection of Fig.14 it can be observed that the maturation level will presumably occur at around \$7000.
2- The left shoulder has occurred after the exponential growth which has been supported by the line $L_5$ as shown in Fig. 6. The beginning level of this growth, which is around $4000, can be taken as the base for the left shoulder if the expected SHS occurs.

3- The duration of that SHS formation is expected to last till around 2020 or longer and thus, it will be the longest for a parabolic behavior ever occurred in the history of the Dow. This means that the index will undergo a downward acceleration for a long time.

As a result, the index may very well test the base levels after crossing down the maturation line. This is because the levels for the maturation and base are close; i.e., $7000 and $4000, respectively. Moreover, the occurrence of such a long-term SHS formation may portend the upcoming of a severe worldwide economic crisis since several SHS formations are maturing also in other world market indices, such as the stock exchange indices of United Kingdom, Germany, France, Hong Kong and the stock exchange indices are expected to be the precursors of economies.

**3. Discussion and conclusion:**

The results of the model can be summarized in the following list.

1- The financial signals can be empirically decomposed into two components as random and non random. Moreover, if the non random component is exploited well, then it can be used for mid-term trading in the stock markets. This is because that component follows one of the simple trends given in Table 1 most of the time.

2- The prices may be in equilibrium only in an epoch for a horizontal behavior. In other words, if a down or up trend have occurred in an epoch, then it means that either buyers or sellers are dominant in that epoch, respectively. This result supports the claim that the exchange markets but particularly the Dow are not in equilibrium. In other words, profitable trading can be made in the exchange markets, or at least in Dow, when the prices are in disequilibrium.

3- What forms a trend may be the collective behavior of the market society under the effect of the past and running news, and expectations about the future. In other words, news may play minor role but only in the short-term. However, the accumulation of bad or good news may be important in the mid- or long-term.

An important argumentation may be about the credibility of the long lasted supporting lines, resisting lines and formations. For example, the following questions can be posed by the reader: Why did the prices not cross over the resisting line $L_2$ of Fig. 3, in the first basic MRT even though the prices approached to or tested the given line several times beforehand? How can the two previous events for the prices $P_1$ and $P_4$ in the same epoch be effective on the future prices, with regard to those two? May it be possible that the importance of a line continues through the history of Dow? How can the line $L_3$ resist against the Dow in 1999 and 2000 in contrast to the evidence that it has supported the same index for many years beforehand? Have those long lasted features been created by the planned behaviors of a group of traders who control big capital or portfolio, or independently of the traders? Similar questions can be listed without limit; however, no answer can be given to them. This is because no snooping is possible for the plans or aims of the traders. However, it is possible to claim that the known shape of the Dow has emerged as a result of the collective behavior of the market community at a time.

The last but not the least important claim may be that more successful technical methods with regard to those concerned in this work could have been advanced. However, if a trading method or rule is known and applied by numerous traders, then it may not work well. As a result, the best working method should be the one which is not publicized.

**ACKNOWLEDGEMENT**
The author thanks to Dietrich Stauffer for his kind helps. This paper is the follow up of[11].

## TABLES

| Name of the MRT | Time equation, $X_n(\tau_n)$ | Speed, $\Delta(\Delta X_n(\tau_n))$ | Acceleration, $\Delta(\Delta X_n(\tau_n))$ |
|---|---|---|---|
| Linear | $\Gamma + \Xi\tau_n$, | $\Xi$ | 0 |
| Parabolic | $\Gamma + \Xi\tau_n + ½\Lambda\tau_n^2$ | $\Xi + \Lambda\tau_n$ | $\Lambda$ |
| Exponential | $\Gamma\exp(\Omega\tau_n)$ | $\Omega X_n(\tau_n)$ | $\Omega^2 X_n(\tau_n)$ |
| Cyclic | $\Gamma\sin(W\tau_n + \Theta)$ | $-W\Gamma\cos(W\tau_n + \Theta)$ | $-W^2 X_n(\tau_n)$ |

**Table 1** Above is brief information for the basic MRTs concerned in this work, where the $\tau_n$ is the time with regard to the ending date of the previous; i.e., the $(n-1)^{th}$ MRT and $\Gamma$, $\Xi$, $\Lambda$ and $\Theta$ are the model parameters.

## FIGURES

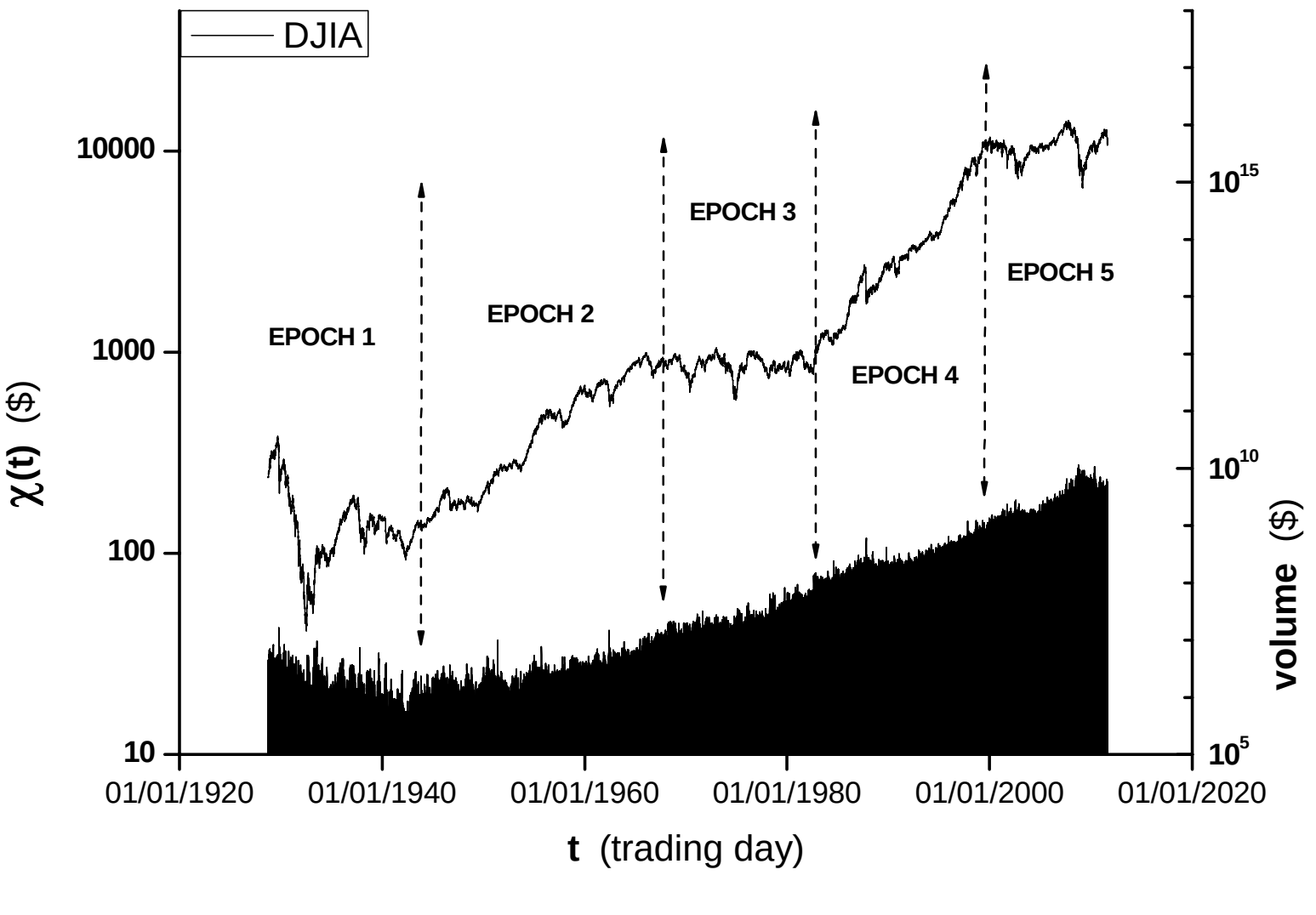

**Figure 1** The daily time course and daily trading volume of the Dow during the basic epochs which are separated by the two headed vertical arrows.

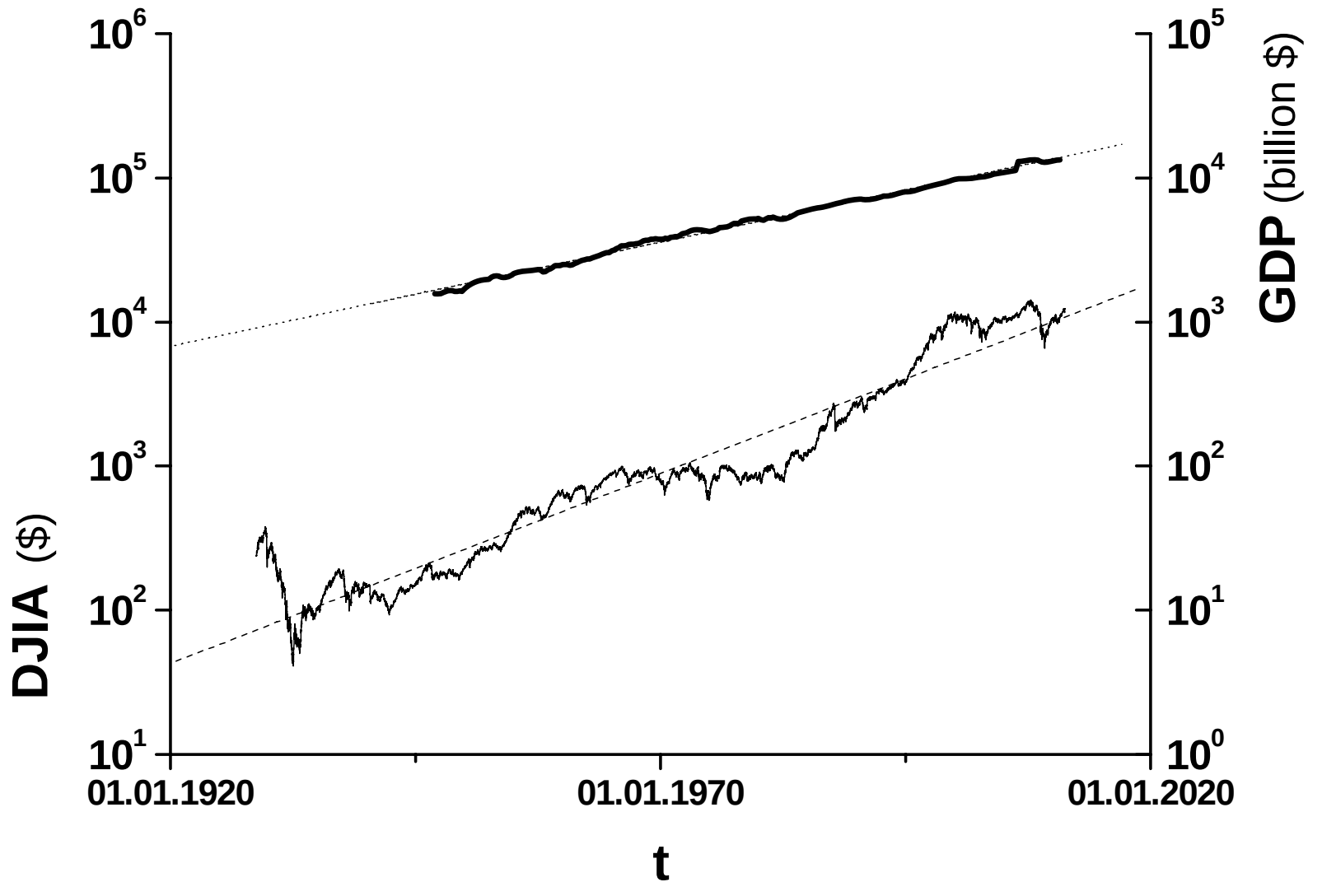

**Figure 2** The gross domestic product (GDP) index of USA (upper part with right vertical axis) and the DJIA index.

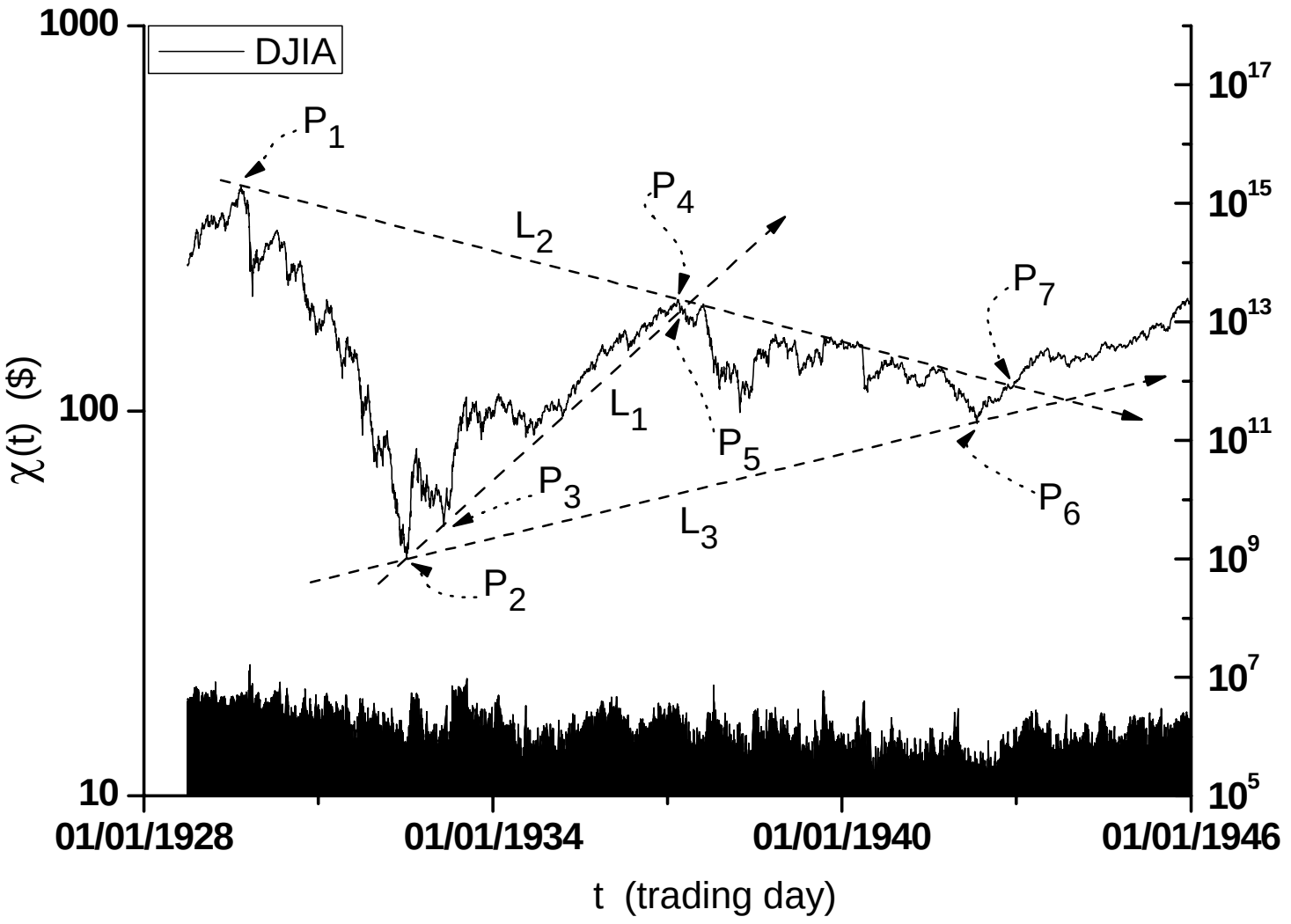

**Figure 3** Behavior of Dow in EPOCH 1.

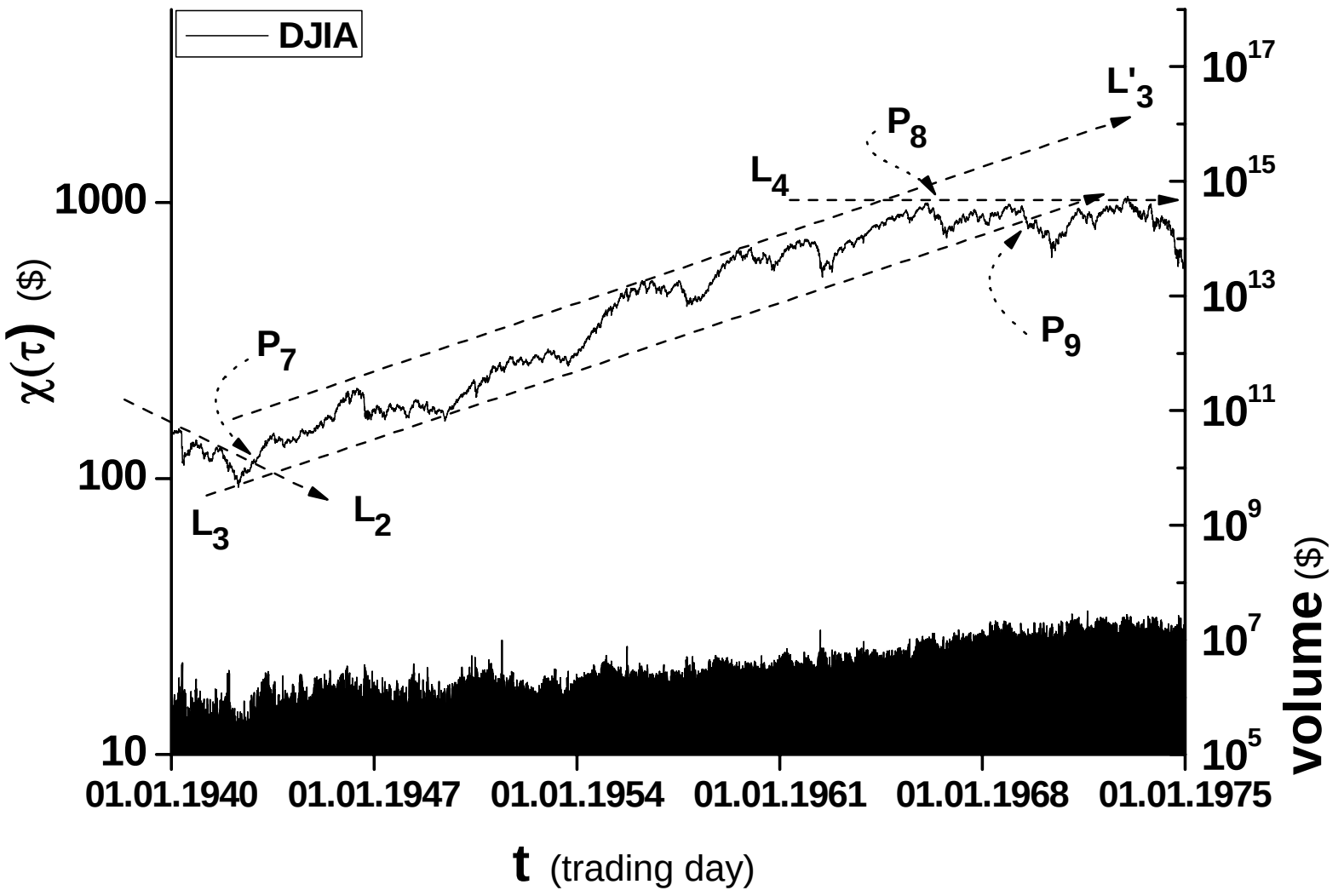

**Figure 4** The exponential growth in the second basic epoch of Dow.

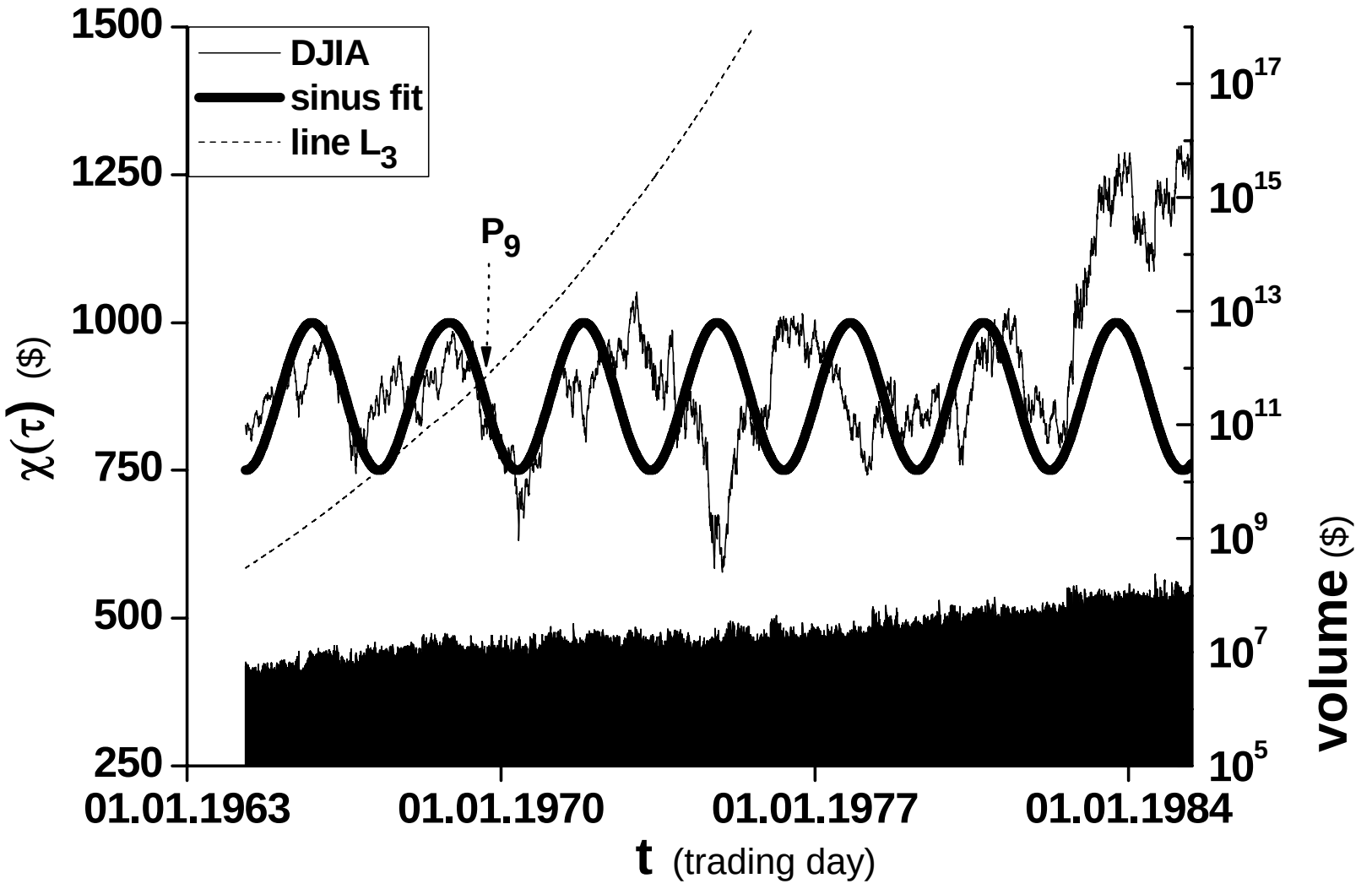

**Figure 5** The nearly cyclic behavior of Dow in the third basic epoch (thin line). See the text around Eq. (11) in Section 2.b for the parameters used in the fitting (thick line).

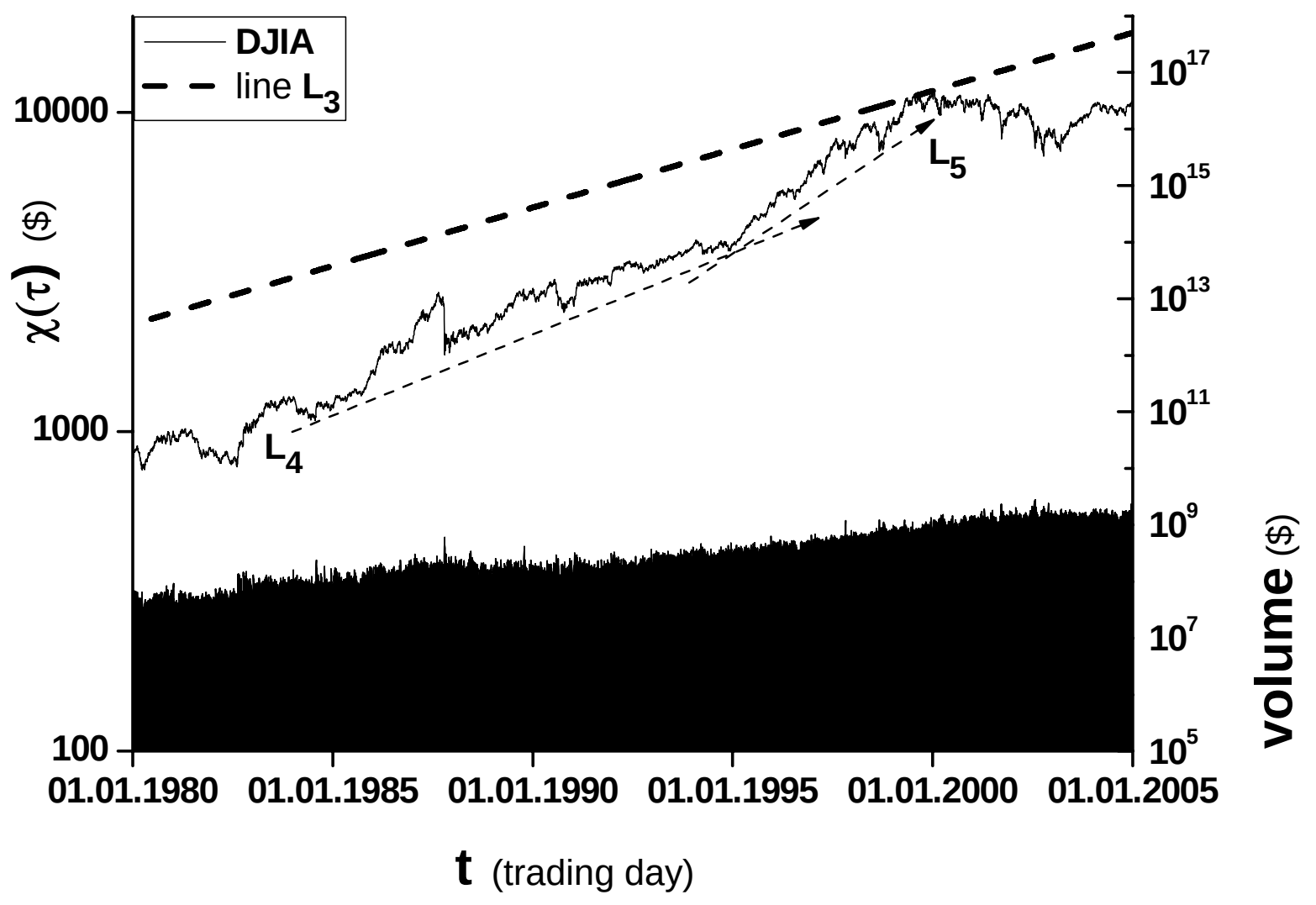

**Figure 6** The exponential behavior of the Dow during the fourth basic epoch.

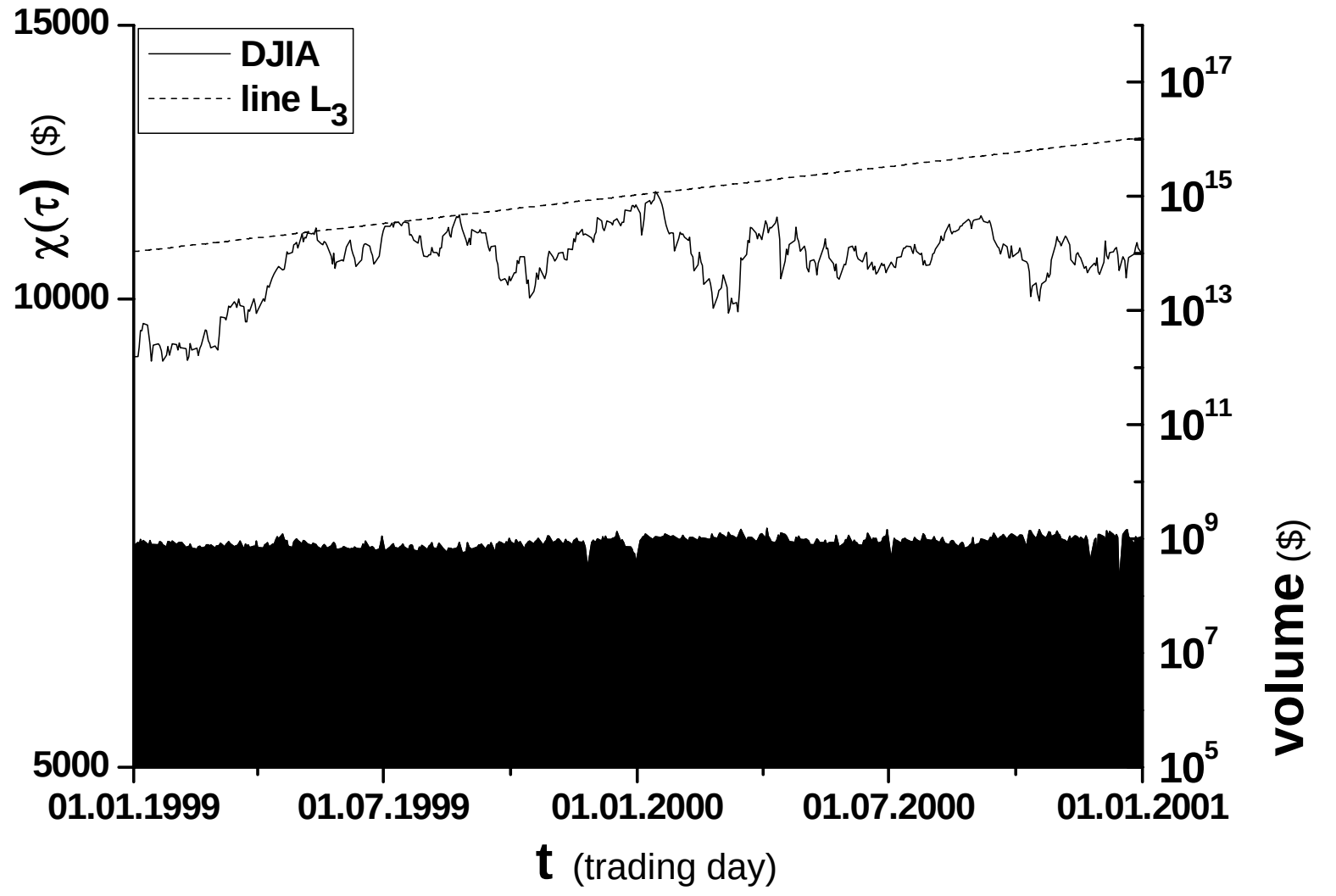

**Figure 7** The behavior of the DJIA index in the years 1999 and 2000.

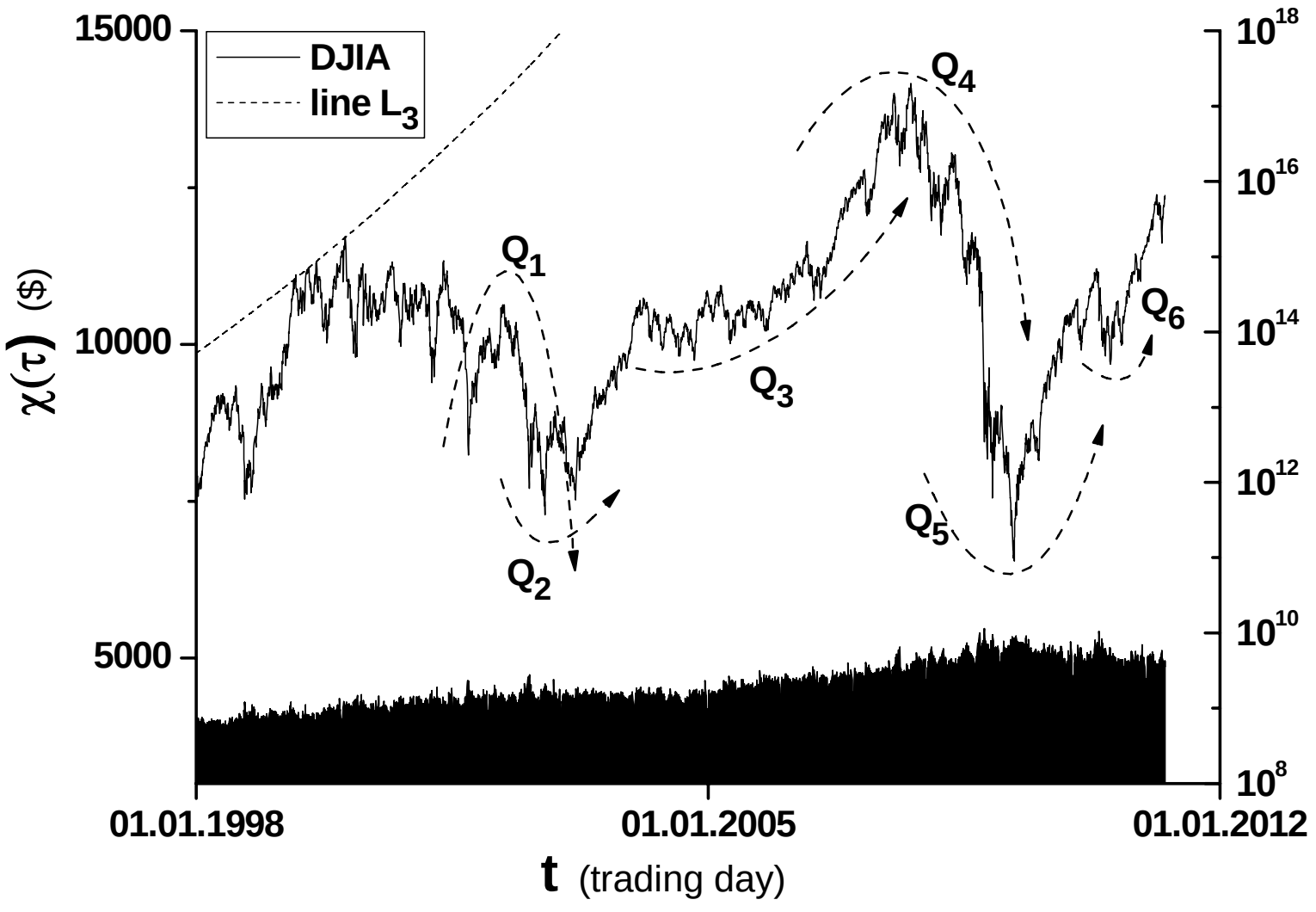

**Figure 8** The behavior of the DJIA index in the current basic epoch where $Q_i$ with $1 \le i \le 6$ indicate the relatively short-term parabolic behaviors.

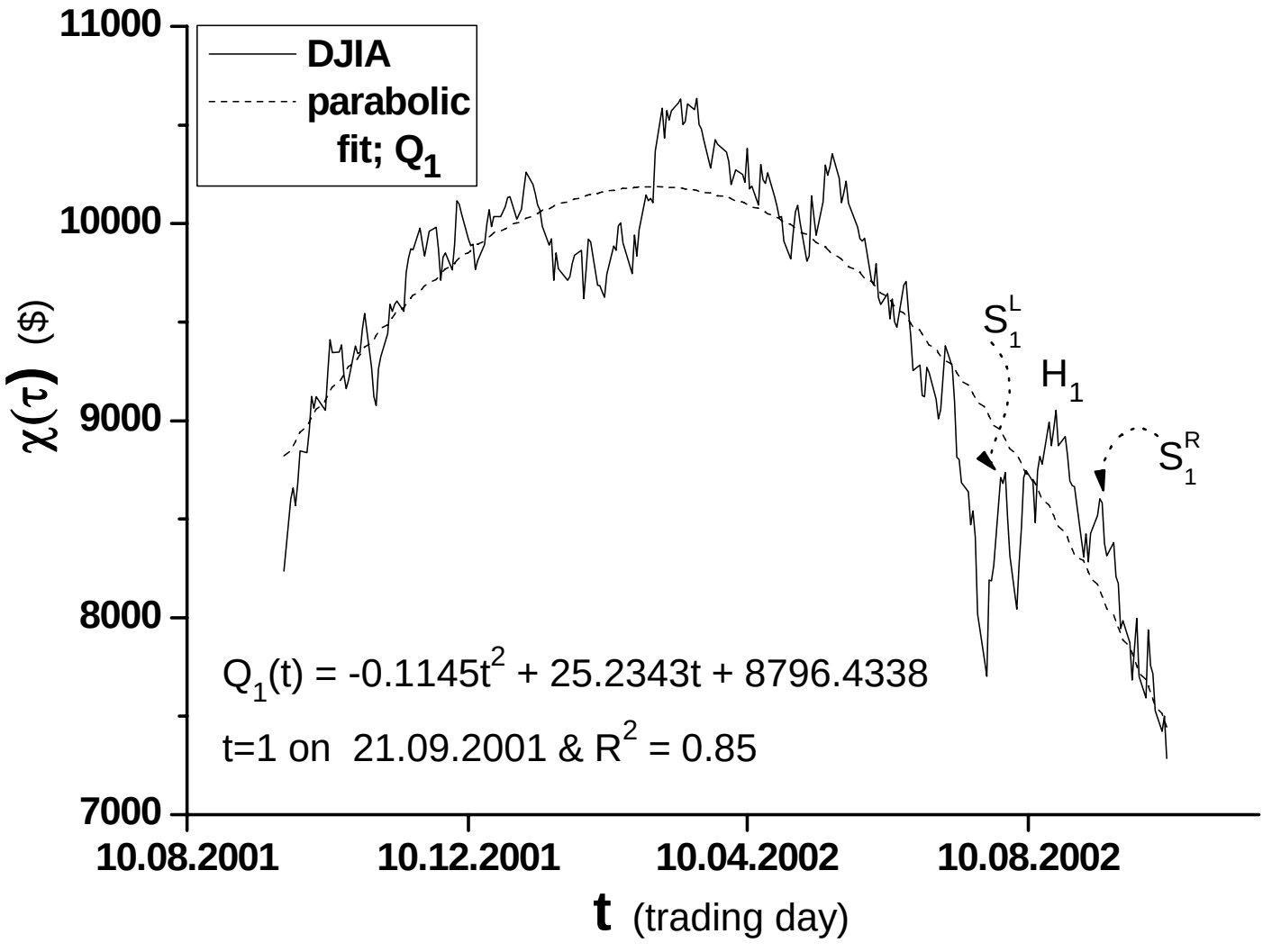

**Figure 9** A parabolic segment of the Dow with a shoulder-head-shoulder formation.

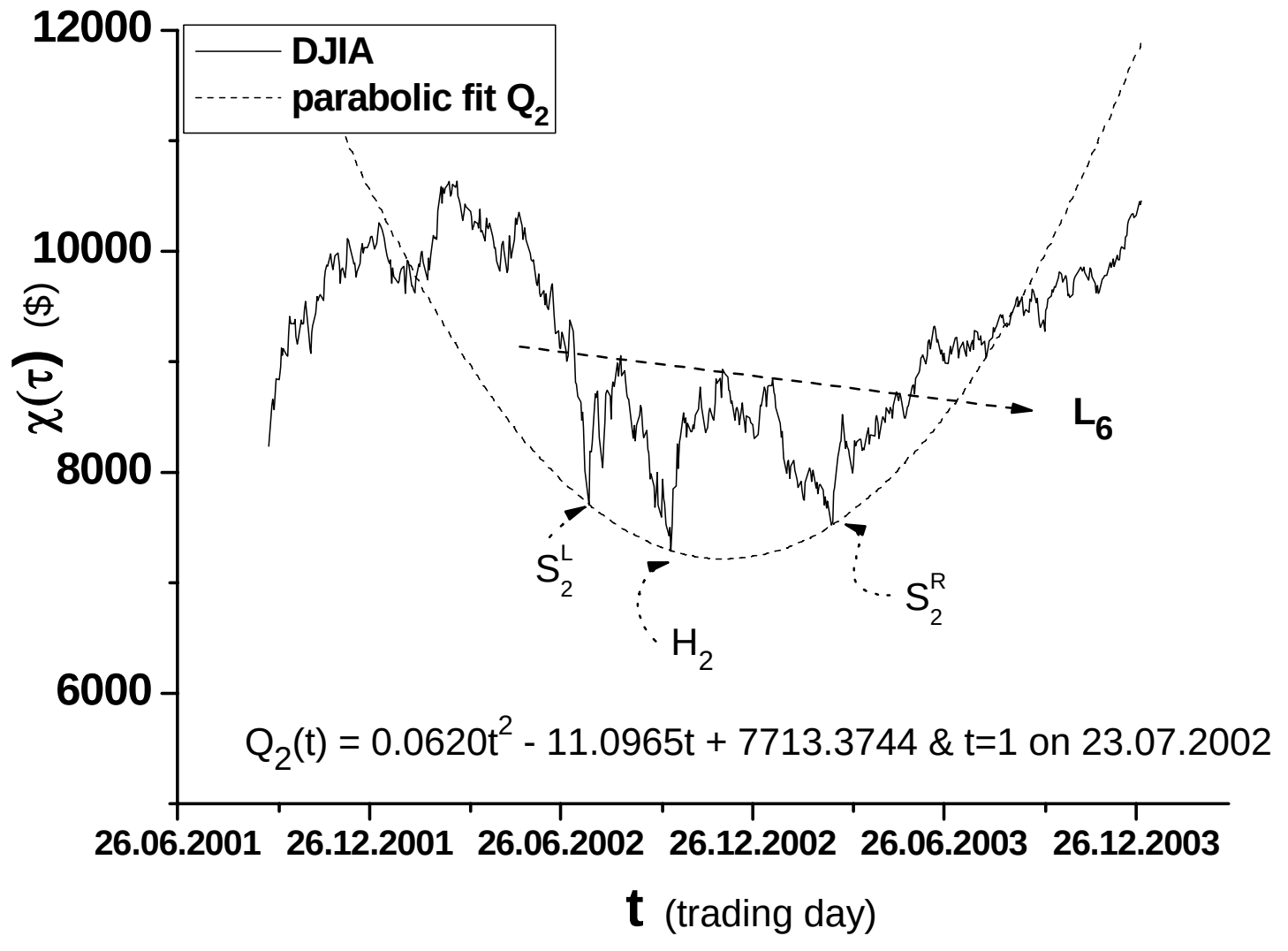

**Figure 10** A parabolic segment of the Dow with a reversed shoulder-head-shoulder formation.

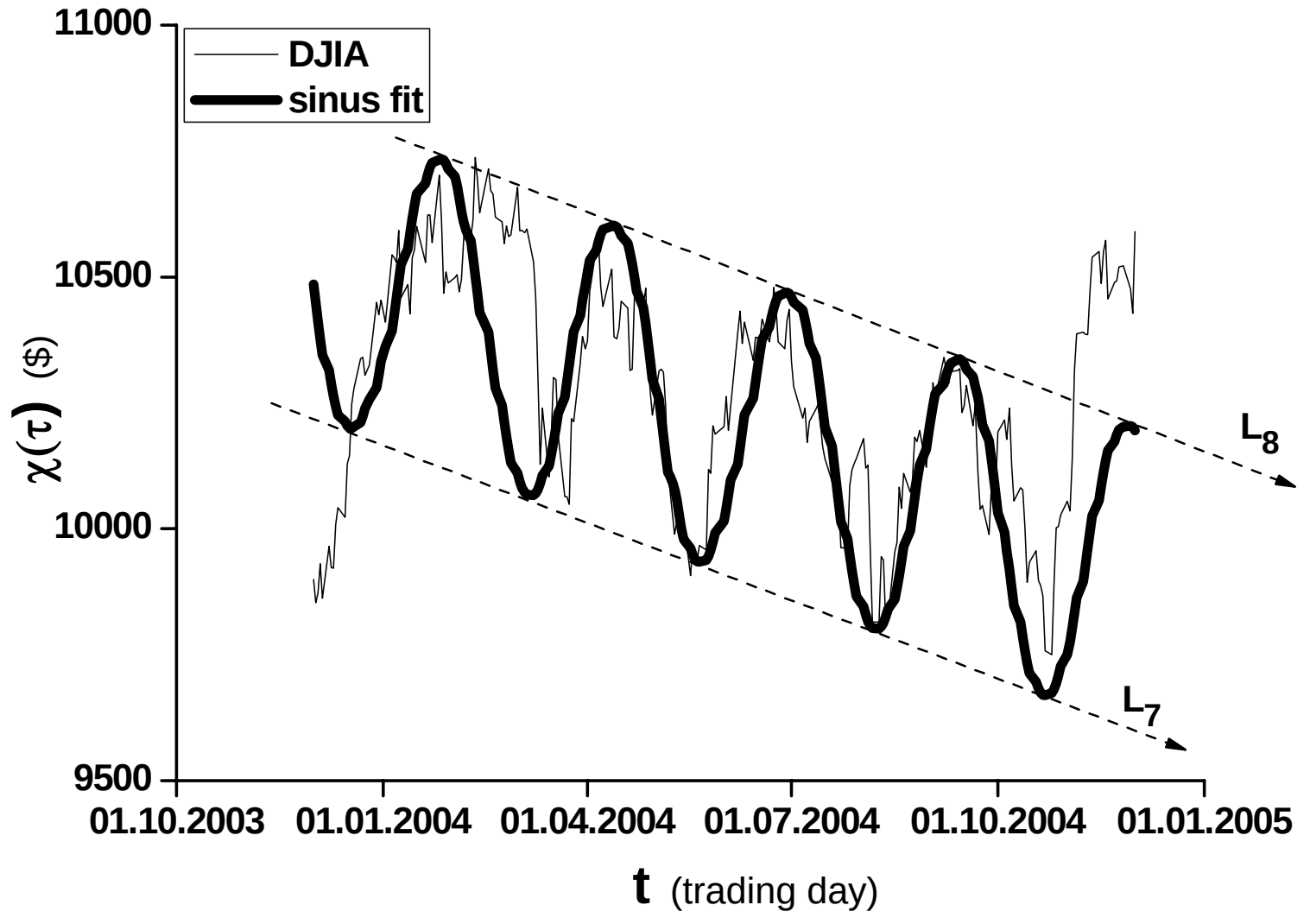

**Figure 11** A cyclic segment of the Dow with a downward average speed.

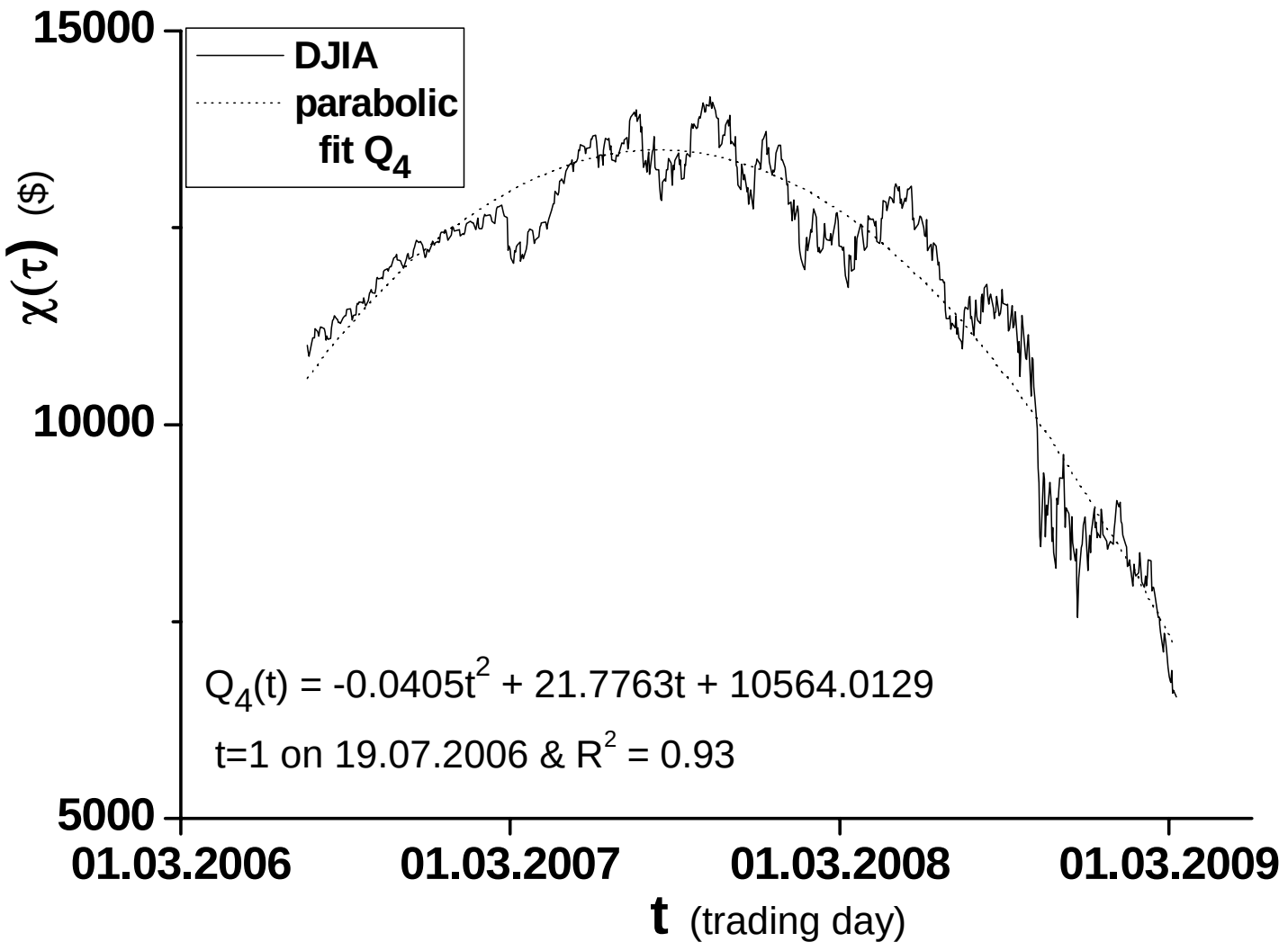

**Figure 12** The behavior of Dow around the historical extreme prices.

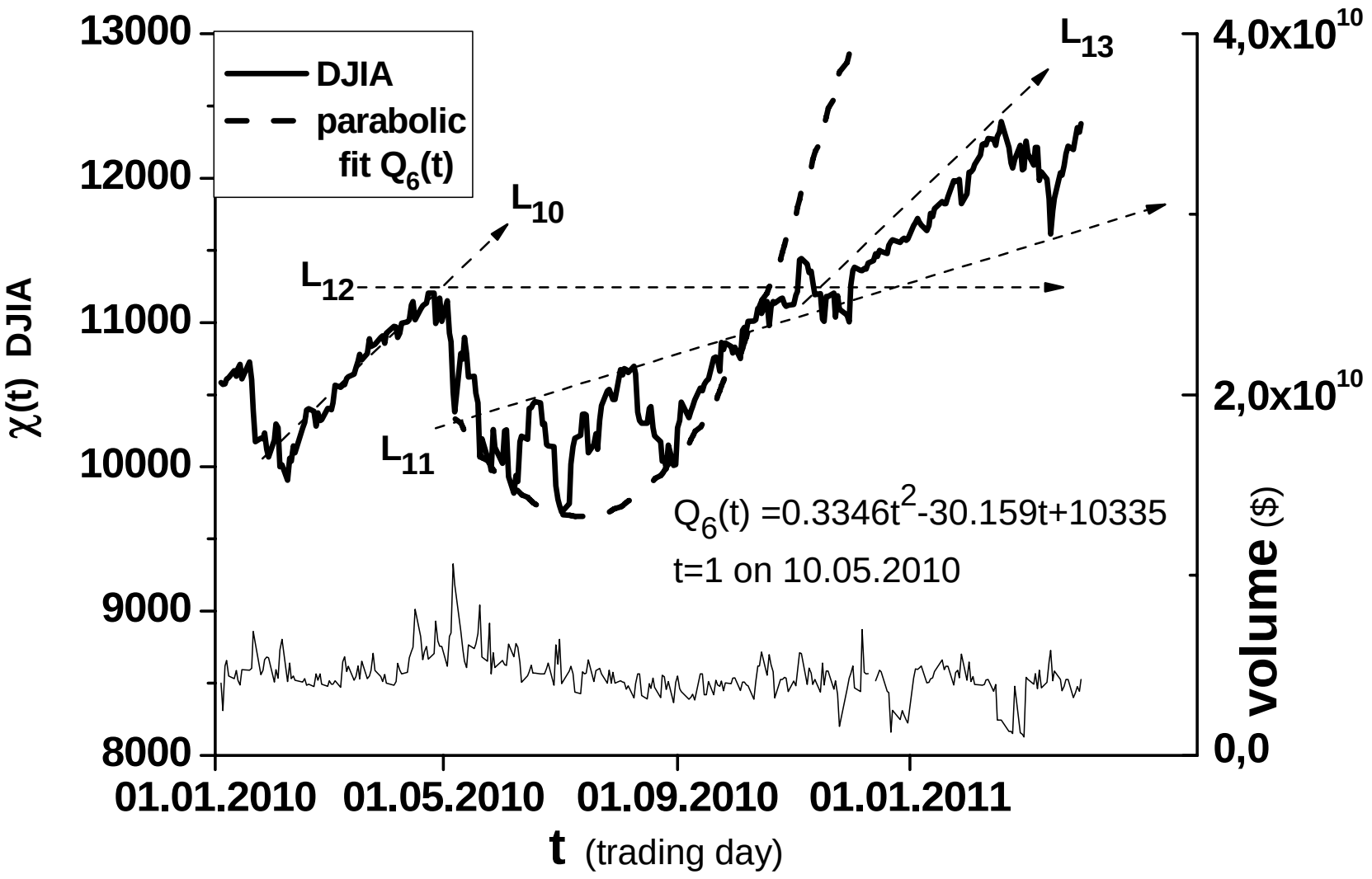

**Figure 13** The behavior of Dow in the time interval between the beginning of 2010 and beginning of the April of 2011.

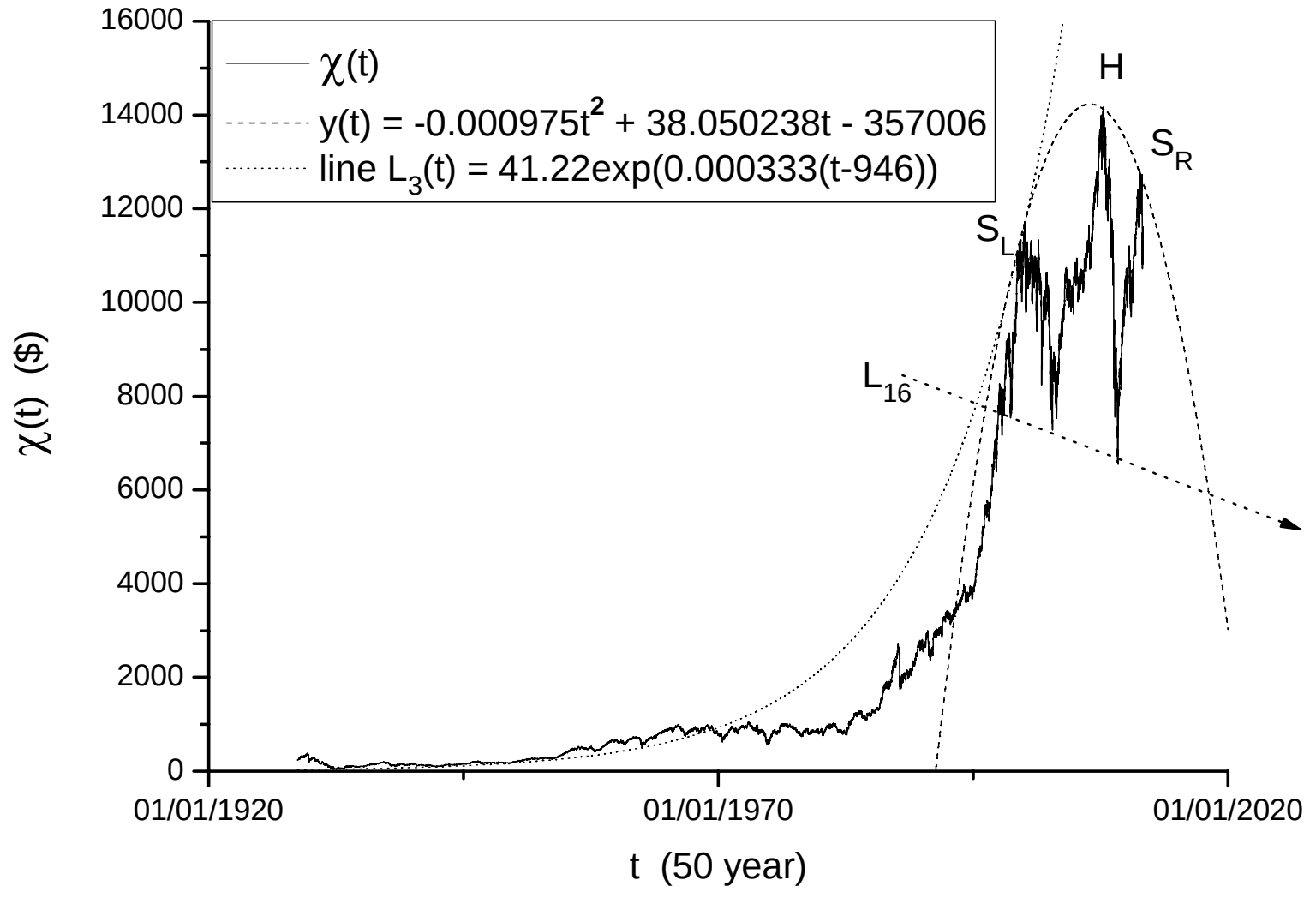

**Figure 14** The current behavior of the Dow.